\documentclass{PoS}
\usepackage{graphicx}

\title{Quark mass dependence of Spin-Orbit force in parity-odd NN system from 2+1 flavor QCD}

\ShortTitle{Quark mass dependence of Spin-Orbit force in parity-odd NN system from 2+1 flavor QCD}
\author{\includegraphics[width=.20\textwidth, bb=0 0 202 118]{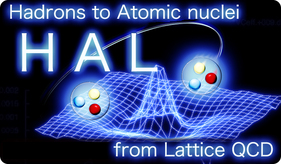}}
\author{\speaker{K. Murano for HAL-QCD Collaboration}
\\
        Yukawa Institute for Theoretical Physics\thanks{YITP-13-125}, Kyoto University, Kitashirakawa Oiwakecho, Sakyo-ku, Kyoto 606-8502, Japan\\
        E-mail: \email{murano@yukawa.kyoto-u.ac.jp}}

      \abstract{We report our recent study of spin-orbit force between
        two nucleons in the parity-odd sector from lattice QCD.
        To examine  the quark mass dependence of  spin-orbit force, we
        construct it from Nambu-Bethe-Salpeter wave
        functions  in  $^3P_0$,  $^3P_1$  and  $^3P_2-^3F_2$  channels
        calculated  from   lattice  QCD  in  the   quark  mass  region
        $m_\pi=702$--$411$ MeV.   The calculation is  performed on Blue
        Gene/Q at  KEK by using $N_f=2+1$  PACS-CS gauge configuration
        generated  by $\mathcal{O}(a)$  improved  Wilson quark  action
        with  RG improved  (Iwasaki) gauge  action. We  find  that the
        potentials  tend   to  become  stronger  as   the  quark  mass
        decreases.}

\FullConference{31st International Symposium on Lattice Field Theory - LATTICE 2013\\
		July 29 - August 3, 2013\\
		Mainz, Germany}

\begin{document}

\section{Introduction}
The study of the nuclear force  is a first step toward the understanding
of the atomic  nuclei beyond the single nucleon.  The nuclear force plays a key role
in describing  various properties of  atomic nuclei and  neutron stars
\cite{Epelbaum:2008ga,Machleidt:2011zz}.
Recently, a  method to extract hadronic interactions  from lattice QCD
has been  proposed, where energy-independent  non-local potentials are
defined  from  the  Schr\"odinger  equation by  using  the  equal-time
Nambu-Bethe-Salpeter   (NBS)    wave   functions   \cite{Ishii:2006ec,
  Aoki:2009ji}.
The method gives  the potentials which are faithful  to the scattering
phase  shift.  This  is  supported  by the  asymptotic  long  distance
behavior of the equal-time NBS wave functions \cite{Aoki:2009ji}.
It has been  successfully applied to the central  and tensor forces in
the    parity-even   NN   system    \cite{Ishii:2006ec,   Aoki:2009ji,
  HALQCD:2012aa}.
It  has been extended  and applied  to various  other systems  such as
hyperon-nucleon   (YN),   hyperon-hyperon   (YY),  meson-baryon,   and
three-nucleons (NNN) \cite{Aoki:2012tk}.
Once these potentials are obtained,  they can be used to study various
physical observables  such as bound states and  scattering phase shift,
by solving the Schr\"odinger equation.
(See Ref.\cite{Kurth:2013tua} for a comparison between the L\"uscher's
method and the potential method for the  $\pi$ $\pi$ scattering phase
shifts, where a good agreement is obtained.)
%
%
%

The  potentials  are  classified  order  by order  in  the  derivative
expansion of the non-local potentials.
%
%
At the leading order (LO),  we have the spin-singlet central potential
$V^{(\pm)}_{\rm  C;  S=0}(r)$,   the  spin-triplet  central  potential
$V^{(\pm)}_{\rm  C;S=1}(r)$ and  the tensor  potential $V^{(\pm)}_{\rm
  T}(r)$, where  the super-index  ``$(\pm)$'' indicates the  parity of
the two-nucleon system.
At  the  next-to-leading  order  (NLO), there  appear  the  spin-orbit
potentials $V^{(\pm)}_{\rm LS}(r)$.
Up   to    the   NLO,   there   are   these    8   independent   local
potentials\cite{Aoki:2009ji}.
So far  our studies  have  been mainly  concentrated on  the
central and the tensor potentials in the parity-even sectors which can
be  obtained  from  the  NBS  wave  functions in  $S$  and  $D$  waves
\cite{Ishii:2006ec,Aoki:2009ji,HALQCD:2012aa}.
%
For complete determination of NN potentials, we need also to determine
the central and  the tensor potentials in the  parity-odd sector at LO
as well  as the spin-orbit (LS)  potentials in both  parity sectors at
NLO.
Especially, the spin-orbit  potential is known to be  important in the
NN  system not  only  to  reproduce the  experimental  phase shift  in
spin-triplet $P$ wave channels, but  also to induce the $P$-wave superfluidity in the stellar environment such as the neutron star interiors
\cite{tamagaki,hoffberg,Baldo:1998ca}.

As our  recent progress, we have  extended our method  to the central,
the tensor and the spin-orbit potentials in the parity-odd sector, and
have    presented    the   first    result    of   these    potentials
\cite{Murano:2013xxa}.  The  calculation was performed by  using the 2
flavor  gauge  configuration  generated  by CP-PACS  collaboration  at
$m_\pi=1133$ MeV  \cite{Ali Khan:2001tx},  where we found  that, while
the qualitative  behavior of resultant potentials  are consistent with
phenomenological  ones,  these potentials  are  still  weak, which  is
considered  to be  caused  by the  heavy  quark mass  employed in  our
simulations.
The main purpose of this paper is to examine the quark mass dependence
of the central, the tensor and the spin-orbit potentials in parity-odd
NN system, by using the $2+1$ flavor gauge configurations generated by
PACS-CS  collaboration  at  $m_\pi=702$,  $570$  and  $411$  MeV
\cite{Aoki:2008sm}.
%
%
%

\section{Definition of the potential}
To   construct  the   NN   potential,  we   consider  the   equal-time
Nambu-Bethe-Salpeter (NBS)  wave function in  the center of  mass (CM)
frame, defined by
\begin{eqnarray}
  \phi_{\alpha,\beta}({\bf r};{\bf k})\equiv \langle 0 | p_\alpha({\bf x})
  n_\beta({\bf y}) | p(+{\bf k})n(-{\bf k})\rangle, \quad ({\bf r}\equiv
  {\bf x}-{\bf y}) \label{eq:NBS}
\end{eqnarray}
where  $p_\alpha({\bf   x})$  and  $n_\beta({\bf   y})$  denote  local
composite nucleon operators with spinor indices $\alpha$, $\beta$,
and  ${\bf k}$  denotes  the asymptotic  momentum  of the  two-nucleon
state.
%
We define NN potentials from the NBS wave function below the inelastic
threshold  ($E <  E_{\rm  th}\equiv 2  m_N  + m_{\pi}$)  by using  the
Schr\"odinger equation \cite{Aoki:2009ji,Aoki:2012tk}
%
\begin{equation}
  \left(
    E_{\bf k} +\frac{\nabla^2}{m_N}
  \right)
  \phi({\bf r};{\bf k})
  =
  \int d^3 r
  \left(
    \mathbb{P}^{(+)}   U^{(+)} ({\bf r}, {\bf r}')
    + \mathbb{P}^{(-)} U^{(-)}({\bf r}, {\bf r}')
  \right)
  \phi({\bf r}';{\bf k}),
  \label{eq:pot1}
\end{equation}
where   $E_{\rm   k}   \equiv   \frac{\bf   k^2}{m_N}$   denotes   the
non-relativistic  energy.   $\mathbb{P}^{(+)}$ and  $\mathbb{P}^{(-)}$
denote  projection operators for  parity-even and  parity-odd sectors,
respectively.
$U^{(+)}$   and  $U^{(-)}$  denote   the  non-local   potentials  for
parity-even and  parity-odd sectors, to which we  apply the derivative
expansion as
\begin{eqnarray}
  U^{(\pm)}({\bf r}, {\bf r}')
  &=&
  V^{(\pm)}({\bf r}, {\bf \nabla})
  \delta({\bf r} - {\bf r}')
  \\\nonumber
  V^{(\pm)}({\bf r},{\bf \nabla})
  &=&
  V^{(\pm)}_{\rm C;S=0}(r)   \mathbb{P}^{(S=0)}
  + V^{(\pm)}_{\rm C;S=1}(r) \mathbb{P}^{(S=1)}
  + V^{(\pm)}_T(r)          S_{12}
  + V^{(\pm)}_{LS}(r)       {\bf L}\cdot{\bf S}
  +({\rm NNLO}),
  \label{eq:pot2}  
\end{eqnarray}
where   $\mathbb{P}^{(S=0)}  \equiv   (1  -   {\bf  \sigma}_1\cdot{\bf
  \sigma}_2)/4$   and    $\mathbb{P}^{(S=1)}   \equiv   (3    +   {\bf
  \sigma}_1\cdot{\bf  \sigma}_2)/4$  denote  the projection  operators
onto  the  total  spin  singlet  and  triplet  sectors,  respectively.
$S_{12}\equiv    3({\bf   r}\cdot{\bf    \sigma}_1)({\bf   r}\cdot{\bf
  \sigma}_2)/{\bf  r}^2   -  {\bf  \sigma}_1\cdot{\bf   \sigma}_2$  is
referred to as  the tensor operator.
${\bf L}\equiv  i\ {\bf  r} \times {\bf  \nabla}$ and ${\bf  S} \equiv
({\bf  \sigma}_1  +  {\bf  \sigma}_2)/2$ denote  the  orbital  angular
momentum operator and the total spin operator, respectively.
%
$V_{\rm C;S=0}^{(\pm)}$,  $V_{\rm C;S=1}^{(\pm)}$, $V_{\rm T}^{(\pm)}$
and $V_{\rm LS}^{(\pm)}$ are  referred to as the spin-singlet central,
the  spin-triplet  central, the  tensor,  and  the spin-orbit  forces,
respectively.
While $V_{\rm  C; S=0}^{(\pm)}$, $V_{\rm C;  S=1}^{(\pm)}$ and $V_{\rm
  T}^{(\pm)}$ are of leading  order(LO) in the derivative expansion of
the   non-local    potential,   $V_{\rm   LS}^{(\pm)}$    appears   at
next-to-leading  order(NLO).  Once  the above  NBS wave  functions are
calculated  in  lattice  QCD  simulations,  these  potentials  can  be
extracted by solving Eq.(\ref{eq:pot1}).

\subsection{Spin-triplet potentials in parity-odd sector
including spin-orbit force}
We restrict ourselves to  the spin-triplet and parity-odd sector.  The
Schr\"odinger equation Eq.(\ref{eq:pot1}) reads
\begin{equation}
  \left(
    E+\frac{\bf \nabla^2}{m_N}
  \right)
  \phi({\bf r})
  =
  \left[
    V_{\rm C;S=1}^{(-)}(r)
    + V_{\rm T}^{(-)}(r) \ S_{12}
    + V_{\rm LS}^{(-)}(r) \ {\bf L}\cdot{\bf S}
  \right]
  \phi({\bf r}).
  \label{eq:SE_S1}
\end{equation}
%
%
In order to determine  $V_{\rm C;S=1}^{(-)}$, $V_{\rm T}^{(-)}$ and
$V_{\rm  LS}^{(-)}$,  we need  three  independent  NBS wave  functions
$\phi_{i}({\bf  r})$ with  $i=1, 2,  3$, for  which we  take  NBS wave
functions      in      $^3P_0(J^P=0^-)$,     $^3P_1(J^P=1^-)$      and
$^3P_2-^3F_2(J^P=2^-)$ channels.
%
$V_{\rm  C;S=1}^{(-)}$, $V_{\rm T}^{(-)}$  and $V_{\rm  LS}^{(-)}$ are
obtained as solutions to Eq.(\ref{eq:SE_S1}) as
\begin{eqnarray}
  \left(
    \begin{array}{c}
      V_{\rm C;S=1}^{(-)}(r) \\
      V_{\rm T}^{(-)}(r) \\
      V_{\rm LS}^{(-)}(r)
    \end{array}
  \right)
  =
  \ M(\vec{r})^{-1}
  \left(
    \begin{array}{c}
      (\nabla^2/m_N + E_1) \ \phi_1({\bf r}) \\
      (\nabla^2/m_N + E_2) \ \phi_2({\bf r}) \\
      (\nabla^2/m_N + E_3) \ \phi_3({\bf r}) 
    \end{array}
  \right), \label{eq:solve}
\end{eqnarray}
where $E_i$  denotes the  non-relativistic energy associated  with the
NBS wave functions $\phi_i({\bf r})$, and $M(\vec{r})$ is a $3\times3$
matrix defined by
\begin{eqnarray}
  M(\vec{r})
  \equiv
  \left(
    \begin{array}{ccc}
      \phi_1({\bf r})  \ \ & S_{12} \phi_1({\bf r}) \ \  & {\bf L}\cdot{\bf S} \phi_1({\bf r}) \\
      \phi_2({\bf r})  \ \ & S_{12} \phi_2({\bf r}) \ \  & {\bf L}\cdot{\bf S} \phi_2({\bf r}) \\
      \phi_3({\bf r})  \ \ & S_{12} \phi_3({\bf r}) \ \  & {\bf L}\cdot{\bf S} \phi_3({\bf r}) \\
    \end{array}
  \right).
\end{eqnarray}


\section{Construction of NBS wave functions by lattice QCD}
The equal-time  NBS wave functions  are obtained from  4-point nucleon
correlation functions on the lattice in the large $t$ region as
\begin{eqnarray}
    G({\bf x}-{\bf y},t-t_0;\mathcal{J}^{J,S=1})
  &\equiv&
  \frac{1}{L^3}
  \sum_{\bf r}
  \langle 0 |
  T\left[
    p({\bf x}+{\bf r},t)
    n({\bf y}+{\bf r},t)
    \mathcal{J}^{J,S=1}(t_0)
  \right]
  | 0 \rangle
  \label{eq:4-point}	
  \\\nonumber
  &\simeq&
  \phi^{J,S=1}_{0}({\bf x}-{\bf y}) \
  a_0
  e^{-E_0 (t-t_0)}, \quad t-t_0 \gg 1,
\end{eqnarray}
where the summation  over ${\bf r}$ in the first  line is performed to
select the  two-nucleon system with vanishing  total spatial momentum.
$\mathcal{J}^{J,S=1}$  denotes a two-nucleon  source operator  for the
total angular momentum $J$ in the spin-triplet parity-odd sector.
For nucleon operators $p(x)$ and $n(y)$, we employ the following local
composite operators
\begin{eqnarray}
  p(x) \equiv \epsilon_{abc} (u_a^T(x) C\gamma_5 d_b(x)) u_c(x), \quad
  n(x) \equiv \epsilon_{abc} (u_a^T(x) C\gamma_5 d_b(x)) d_c(x),
\end{eqnarray}
where $a$, $b$ and $c$ denote color indices.
%
$\phi_0^{J,S=1}({\bf r})$ and $E_0$  denotes the NBS wave function and
the energy  of the ground state  in the total angular  momentum $J$ in
the spin-triplet parity-odd sector, respectively.
The coefficient $a_0     \equiv    \langle     p(+{\bf     k}_0)n(-{\bf    k}_0)     |
\mathcal{J}^{J,S=1}(0)|0\rangle$ is the  overlap factor between states
created by the source and the ground state in this system
with the asymptotic momentum ${\bf k}_0$.

For  the two-nucleon  source operator  $\mathcal{J}^{J,S=1}$,  we take
two-nucleon momentum wall source operator defined by
\begin{equation}
  \mathcal{J}_{\alpha\beta}(f) \equiv \bar P_{\alpha}(f) \bar N_{\beta}(f^*),
\end{equation}
where
\begin{eqnarray}
  \bar P_{\alpha}(f)
  &\equiv&
  \sum_{{\bf x}_1,{\bf x}_2}
  \epsilon_{abc}
  \left(
    \bar{u}_{a}({\bf x}_1) C\gamma_5 \bar{d}_{b}^T({\bf x}_2)
  \right)
  \sum_{{\bf x}_3}\bar{u}_{c,\alpha}({\bf x}_3) f({\bf x}_3)
  \\\nonumber
  \bar N_{\beta}(f)
  &\equiv&
  \sum_{{\bf x}_1,{\bf x}_2}
  \epsilon_{abc}
  \left(
    \bar{u}_{a}({\bf x}_1) C\gamma_5 \bar{d}_{b}^T({\bf x}_2)
  \right)
  \sum_{{\bf x}_3}\bar{d}_{c,\beta}({\bf x}_3) f({\bf x}_3)
\end{eqnarray}
with $f$  being one of the  following source functions,  each of which
corresponds  to a  plane  wave  parallel or  anti-parallel  to one  of
the spatial coordinate axes as
\begin{eqnarray}
 f^{(0)}({\bf r}) \equiv \exp(-2\pi i x/L),
 \quad
 f^{(1)}({\bf r}) \equiv \exp(-2\pi i y/L),
 \quad
 f^{(2)}({\bf r}) \equiv \exp(-2\pi i z/L),
 \nonumber \\
 f^{(3)}({\bf r}) \equiv \exp(+2\pi i x/L),
 \quad
 f^{(4)}({\bf r}) \equiv \exp(+2\pi i y/L),
 \quad
 f^{(5)}({\bf r}) \equiv \exp(+2\pi i z/L). 
\end{eqnarray}
Note that an element $g$ of  the cubic group $O$ with 24 elements acts
on these plane waves as
\begin{eqnarray}
 f^{(i)} \mapsto  \sum_{j}  U_{ij}(g)  f^{(j)},
\end{eqnarray}  
where $U(g)$ is a $6\times 6$ permutation matrix, which servers as the
representation matrix of $g \in O$.
By  a cubic group  analysis, the  orbital part  of this  momentum wall
source is decomposed into $A_1^+ \oplus E^+ \oplus T_1^-$.  Therefore,
for  the parity-odd  sector,  we can  access  $J^P =  (L=T_1^-)\otimes
(S=A_1)=T_1^-  (\simeq 1^-)$ for  the spin-singlet  sector and  $J^P =
(L=T_1^-) \otimes (S=T_1) = A_1^- \oplus E^- \oplus T_1^- \oplus T_2^-
(\simeq  0^-\oplus 2^- \oplus  1^- \oplus  2^-)$ for  the spin-triplet
sector.

The  momentum  wall source  
with  a  definite total  angular
momentum is now constructed as
\begin{equation}
  \mathcal{J}^{J}_{\alpha\beta}(f^{(i)})
  \equiv
  \frac{d^{(J)}}{24}
  \sum_{g\in O}
  \chi^{(J)}(g^{-1})
  U_{ij}(g)
  \mathcal{J}_{\alpha'\beta'}(f^{(j)})
  S^{-1}_{\alpha'\alpha}(g^{-1})
  S^{-1}_{\beta'\beta}  (g^{-1})
\end{equation}
where  $d^{(J)}$  and $\chi^{(J)}(g)$  denote  the  dimension and  the
character for the irreducible representation $J$ of the cubic group,
respectively.  
Hereafter the  Dirac indices $\alpha,  \beta$ are restricted  to upper
components (in the Dirac representation).
The  total spin  $S$  is  projected by  the  spin projection  operator
$\mathbb{P}^{(S)}$ as
\begin{equation}
  \mathcal{J}^{J,S}_{\alpha\beta}(f^{(i)})
  \equiv
  \mathbb{P}^{(S)}_{\alpha\beta,\gamma\delta}
  \mathcal{J}^{J}_{\gamma\delta}(f^{(i)}).
\end{equation}
%
Finally, the parity projection is performed by
\begin{equation}
  \mathbb{P}^{(\pm)}
  \mathcal{J}_{\alpha\beta}(f^{(i)})
  \equiv
  \frac1{2}
  \left(
  \mathcal{J}_{\alpha\beta}(f^{(i)})
  \pm
  \mathcal{J}_{\alpha\beta}(f^{(i)*})    
  \right).
\end{equation}
Note that $f^{(i)*}$  is the plane wave with  the opposite momentum of
$f^{(i)}$.  (For  detail of  the construction of  two-nucleon momentum
wall source operator, see Ref.\cite{Murano:2013xxa})


\section{Numerical results}
Our  calculation  is  performed  by  using $N_f=2+1$  full  QCD  gauge
configurations generated by PACS-CS collaboration on a $32^3\times 64$
lattice\cite{Aoki:2008sm},  which  employs   the  RG  improved  action
(Iwasaki  action)  at  $\beta=1.90$  leading to  the  
lattice  spacing $a^{-1}  =  2.176(31)$ GeV  ($a  =  0.0907(13)$  fm) and  the  lattice
extension $L \simeq 2.9$ fm.
As the  quark action, it employs the  $\mathcal{O}(a)$ improved Wilson
quark   action   (clover    action)   with   $C_{\rm   SW}=1.715$   at
$\kappa_{ud}=0.13700$,  $0.13727$ and $0.13754$ and $\kappa_{s}=0.13640$.
These $\kappa_{ud}$ correspond to the pion mass $m_\pi=702(1)(10)$, $570(2)(8)$
and  $411.3(2)(6)$  MeV and  the  nucleon  mass  $m_N=1538(5)(23)$, $1411(12)(20)$  and
$1215(12)(17)$ MeV, respectively. The first errors are statistical and the second ones are the systematic errors coming from ambiguity of
 the lattice scale. 
The  4-point  nucleon  correlation functions  Eq.(\ref{eq:4-point})  are
calculated with  the periodic  and Dirichlet boundary  conditions along
the spatial and the temporal directions, respectively.  
To improve the statistics, we use 4 source points by temporally shifting
the gauge configurations.
The charge conjugation and time reversal symmetries are used to double
the number of statistical data.

We calculate the NBS  wave functions with $J^P=A_1^-,T_1^-$ and $E^-$,
whose   dominant  components  correspond   to  $^3P_0$,   $^3P_1$  and
$^3P_2-^3F_2$, respectively.
In order  to extract the  potentials, we solve  Eq.(\ref{eq:SE_S1})
%
by  using the time-dependent method\cite{HALQCD:2012aa}
which  can efficiently  extract the  potentials without  requiring the
ground state  saturation of the 4-point  nucleon correlation functions
Eq.(\ref{eq:4-point}).

\begin{figure}[h]
\begin{center}
\scalebox{0.5}{\includegraphics[]{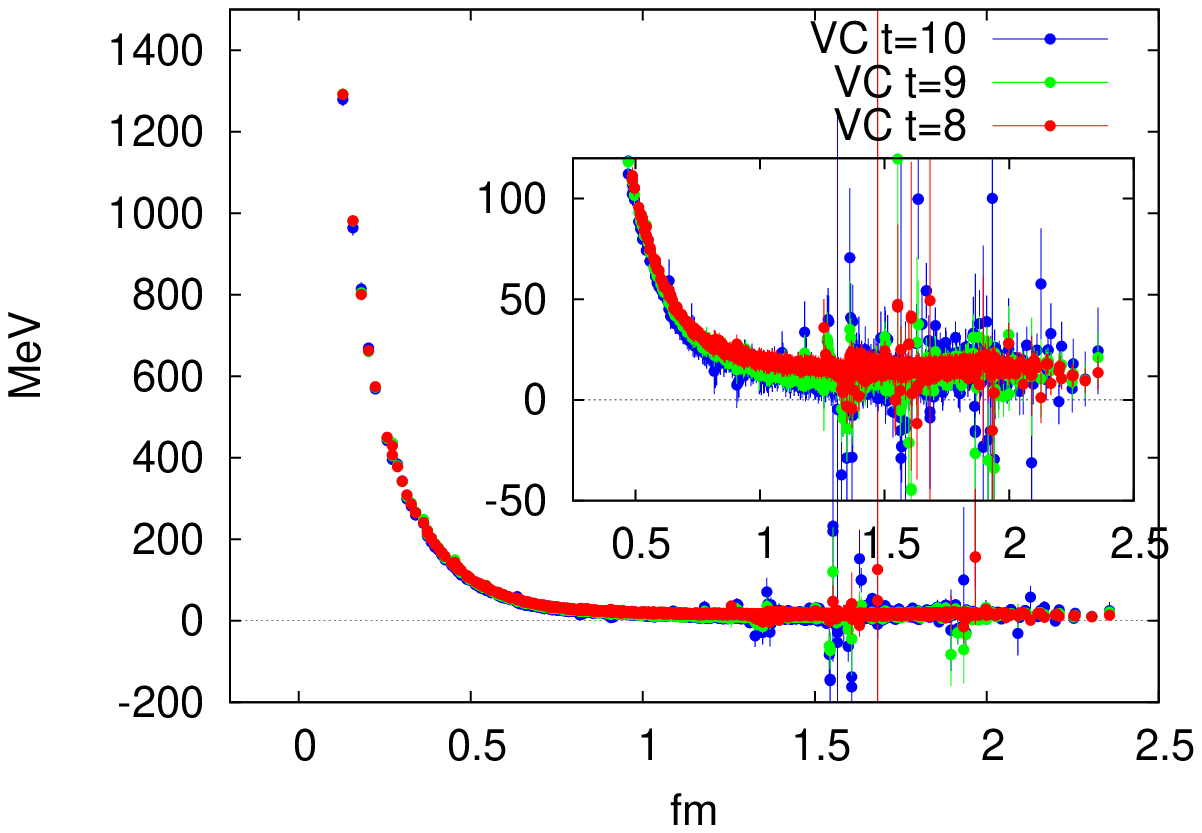}}
\scalebox{0.5}{\includegraphics[]{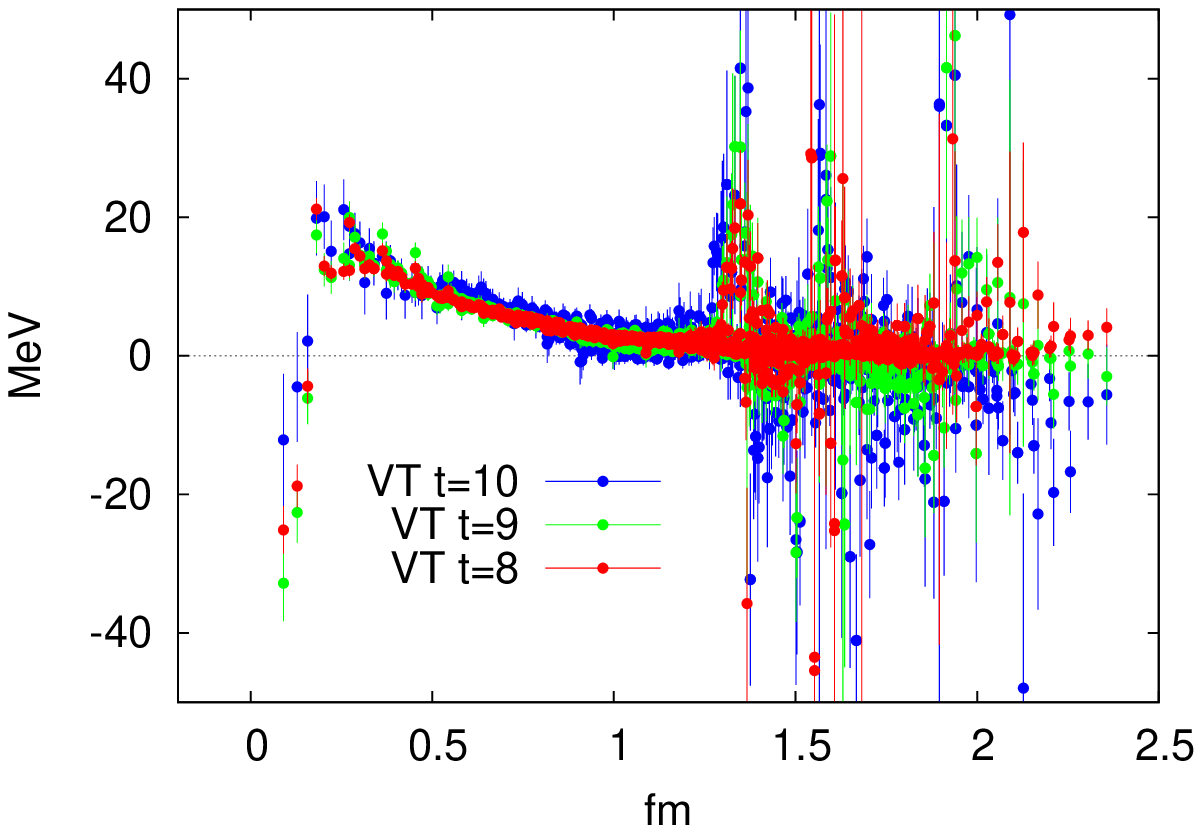}}
\scalebox{0.5}{\includegraphics[]{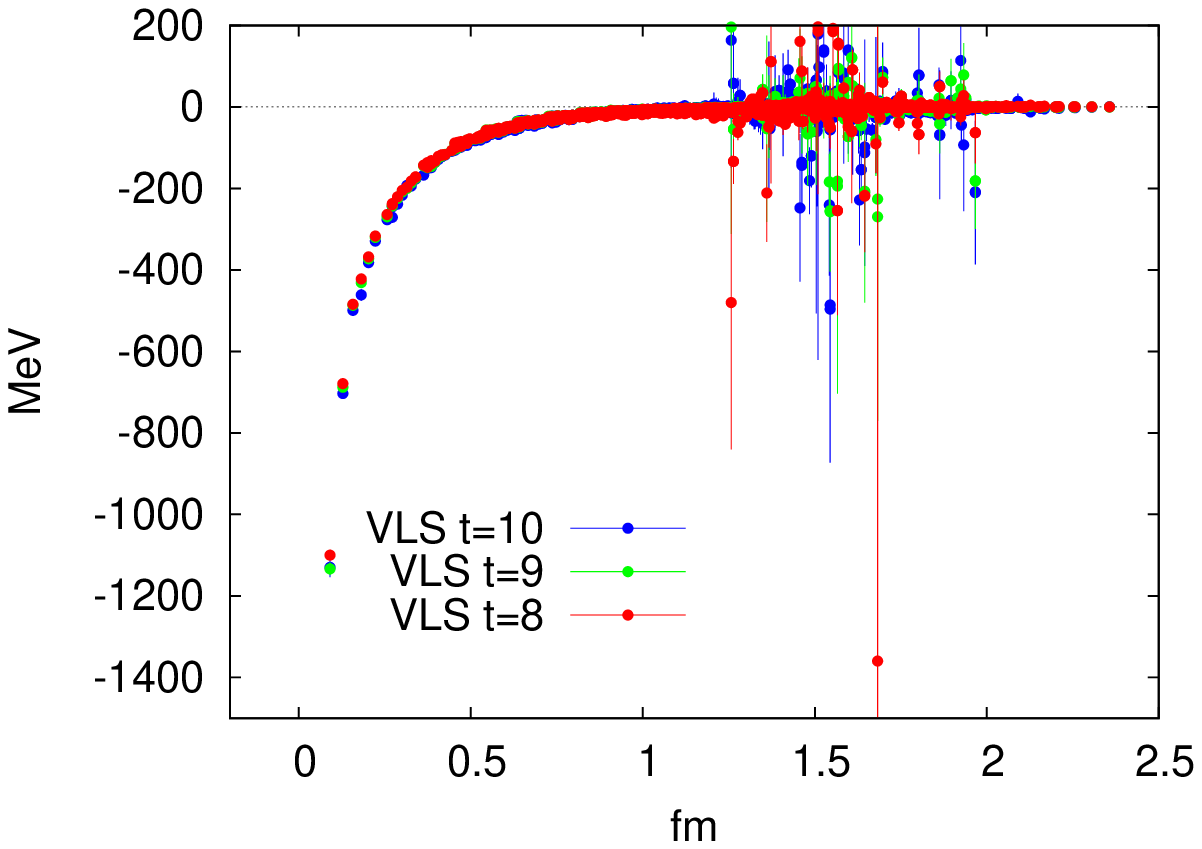}}
\vspace{-0.5cm}
\caption{The $t$-dependence of central(top left), tensor(top right) and spin-orbit(bottom) potentials in the spin-triplet and
  parity odd sector for  $t-t_0=8$(red), $9$(green) and $10$(blue) for
 pion masses  $702$ MeV.}
\label{fig:t-dep}
\end{center}
\end{figure}
Preliminary  results of  the central,  the tensor  and  the spin-orbit
potentials at  $t-t_0=8, 9, 10$  for $m_\pi=702$ MeV are  presented in
Fig.\ref{fig:t-dep}.
We see similar behaviors observed in Ref.\cite{Murano:2013xxa} such as
(1) the central potential has  a repulsive core at short distance, (2)
the  tensor potential  is weak  and positive,  and (3)  the spin-orbit
potential is negative and strong.
However, we observe that the  convergence of long distance part of the
central potential is very slow,
which may be caused by the NNLO terms in the derivative expansion of
the non-local potential or the inelastic contribution in the 4-point
nucleon correlation function Eq.(\ref{eq:4-point}).
Similar tendency is seen for the case of other quark masses.
To achieve the time slice saturation, we need to use the NBS wave
functions at somewhat larger $t$.
\begin{figure}[h]
\begin{center}
\scalebox{0.48}{\includegraphics[]{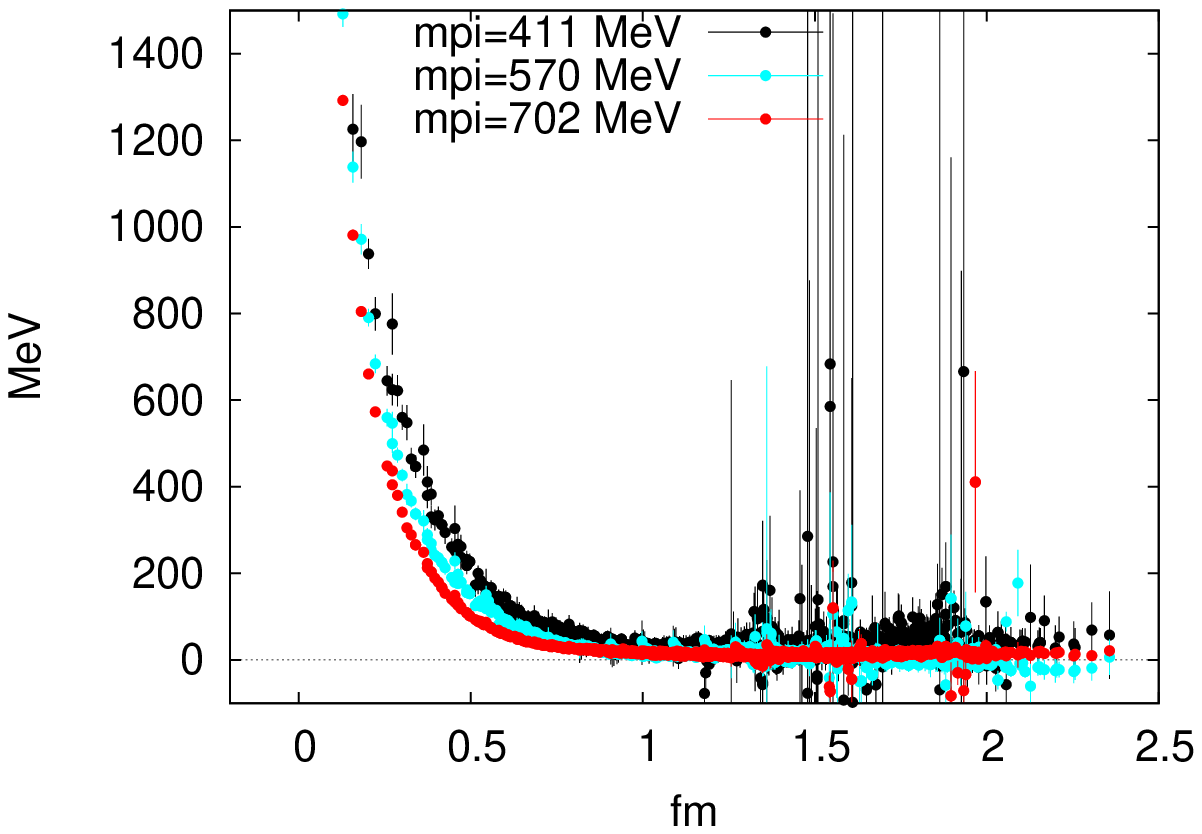}}
\scalebox{0.48}{\includegraphics[]{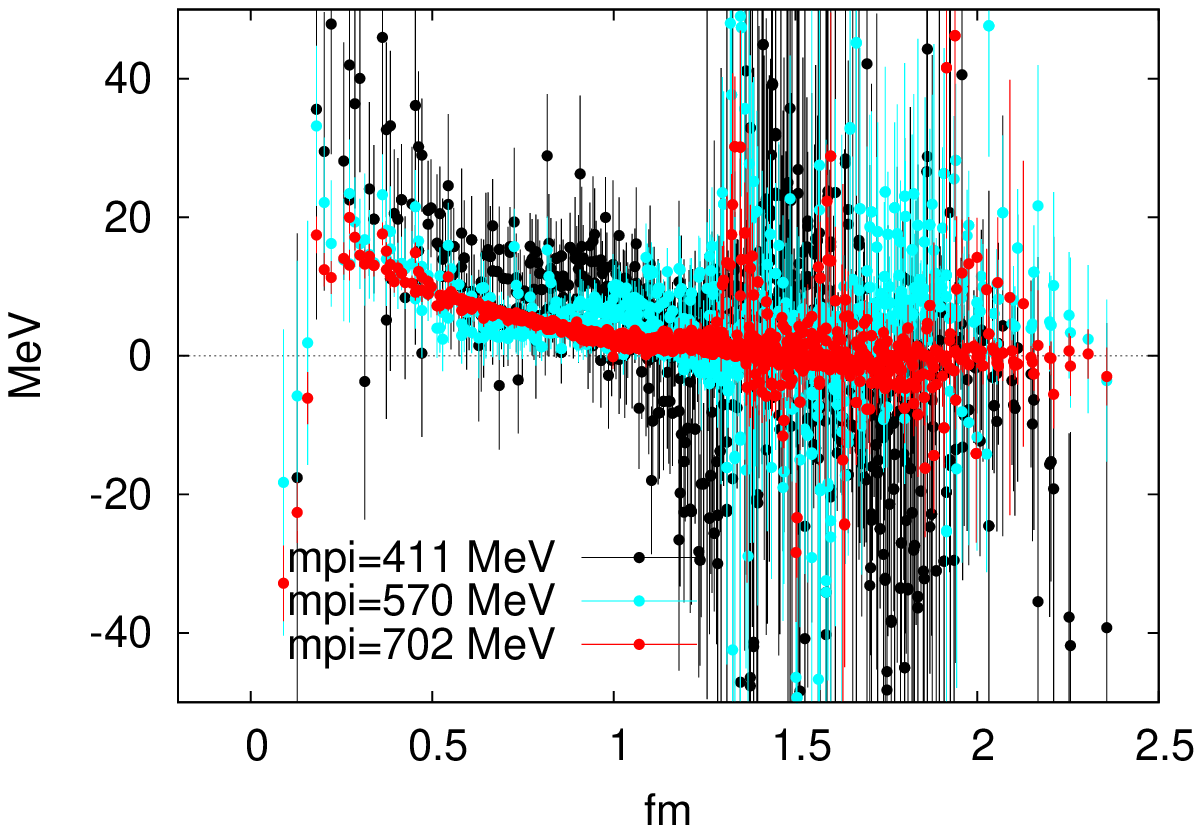}}
\scalebox{0.48}{\includegraphics[]{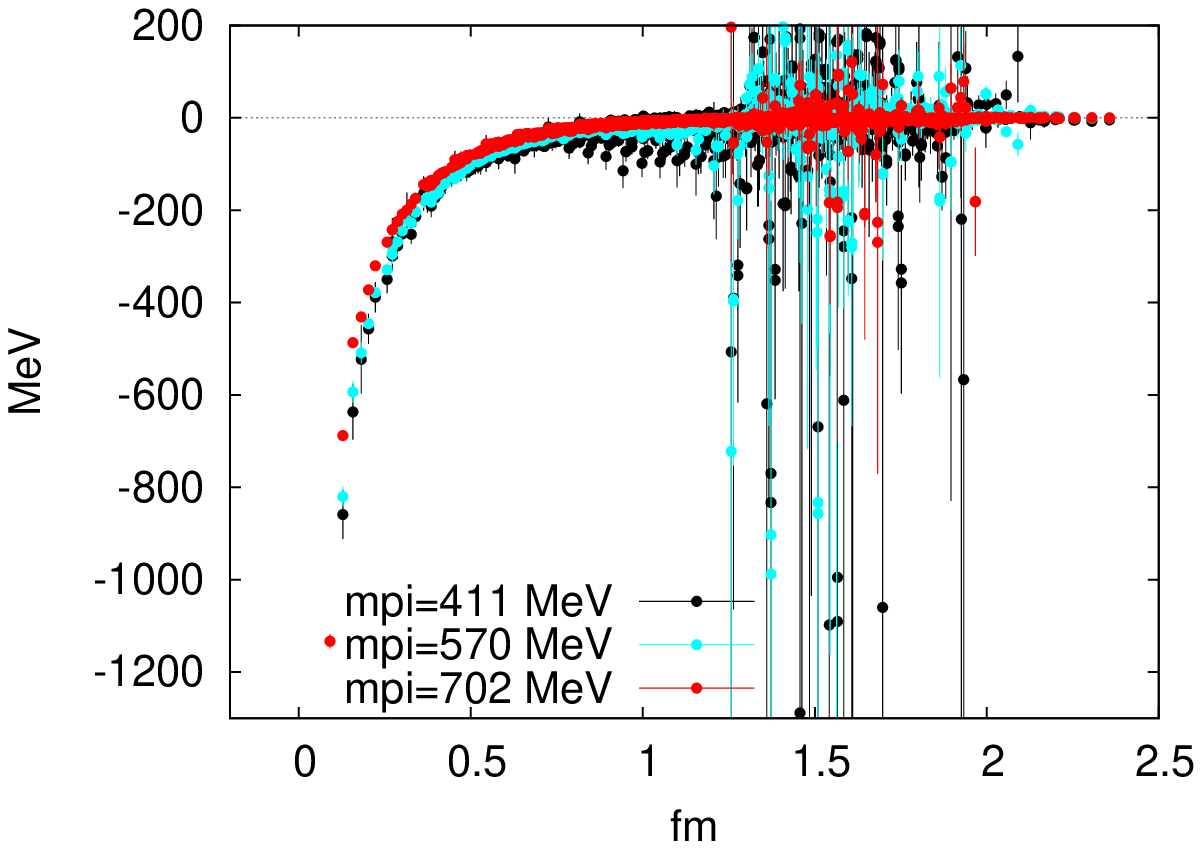}}
\end{center}
\vspace{-0.5cm}
\caption{The quark  mass dependence  of the central  potential (left),
  tensor potential  (right) and  spin-orbit potential (bottom)  in the
  spin-triplet and parity-odd sector.  Red, blue and black points
  corresponding to $m_\pi=702$, $570$ and $411$ MeV, respectively.  }
\label{fig:V}
\end{figure}
Preliminary results on  the quark mass dependence of  the central, the
tensor,  and  the spin-orbit  potentials  at  $t-t_0=9$  are shown  in
Fig.\ref{fig:V}.  Red, blue and black data correspond to the pion mass
$m_\pi=702$, $570$ and $411$ MeV, respectively.
We  find the  tendency that  these potentials  become stronger  as the
quark mass decreases.
In order to discuss the behaviors of  the P-wave phase shifts,
it is necessary to achieve the time slice saturation of the central potential.
For this, it is important to increase the statistics.
\section{Summary}
%
As a continuation  of our previous studies of  the central, the tensor
and  the  spin-orbit  potential  in  the parity-odd  sector,  we  have
examined  a quark  mass dependence  of these  potentials by  using 2+1
flavor   gauge   configurations  which   are   generated  by   PACS-CS
collaboration at $m_\pi=702$, $570$ and $411$ MeV.  Although the
time  slice saturation  is  not  achieved yet,  we  have observed  the
tendency  that these  potentials  become stronger  as  the quark  mass
decreases.

The lattice QCD  calculation has been done on Blue  Gene/Q at KEK under
the support of the  Large Scale Simulation Program No.12/13-19(FY2013)
and No.12-11(FY2012) of  High Energy Accelerator Research Organization
(KEK).
We  are grateful  for authors  and maintainers  of  CPS++\cite{cps}, a
modified  version  of which  is  used  for  simulations done  in  this
report.
We  thank  PACS-CS   collaboration  \cite{Aoki:2008sm}  and  ILDG/JLDG
\cite{jldg} for 2+1 flavor QCD gauge configurations.
This research is supported in part by MEXT Grant-in-Aid for Scientific
Research  (No.25287046), for Scientific  Research on  Innovative Areas
(No.2004:  20105001,  20105003)   and  SPIRE  (Strategic  Program  for
Innovative REsearch).

\end{document}